# 3D printable multimaterial cellular auxetics with tunable stiffness

Krishna Kumar Saxena 1[a,b,*], Raj Das 2[a,c], Emilio P. Calius 3[b]

[a] Department of Mechanical Engineering, Centre for Advanced Composite Materials, University of Auckland, 20 Symonds Street, Auckland 1010, New Zealand.
[b] Future material and structures/ICT research group, Callaghan Innovation, Industrial Research Labs, Auckland, New Zealand.
[c] Sir Lawrence Wackett Aerospace Research Centre, School of Engineering, RMIT University, GPO Box 2476, Melbourne, VIC 3001, Australia
[*] Current affiliation: Department of Mechanical Engineering, KU Leuven, Belgium.

[1]ksax995@aucklanduni.ac.nz

Abstract
Auxetic materials are a novel class of mechanical metamaterials which exhibit an interesting property of negative Poisson's ratio by virtue of their architecture rather than composition. It has been well established that a wide range of negative Poisson's ratio can be obtained by varying the geometry and architecture of the cellular materials. However, the limited range of stiffness values obtained from a given geometry restricts their applications. Research trials have revealed that multi-material cellular designs have the capability to generate range of stiffness values as per the requirement of application. With the advancements in 3D printing, multi-material cellular designs can be realized in practice. In this work, multi-material cellular designs are investigated using finite element method. It was observed that introduction of material gradient/distribution in the cell provides a means to tune cellular stiffness as per the specific requirement. These results will aid in the design of wearable auxetic impact protection devices which rely on stiffness gradients and variable auxeticity.

## 1 INTRODUCTION

Materials with positive Poisson's ratio are known to the society and even zero Poisson's ratio materials such as cork are common in everyday life. However, negative Poisson's ratio materials do exist. Some instances of auxetic behavior in nature can be observed but they are very rare. Negative values of Poisson's ratio can be achieved in man-made materials. Auxetic materials (Evans and Alderson 2000), which can be considered to be a class of mechanical metamaterials, are characterized by having a negative value of Poisson's ratio. They get shorter in transverse direction when compressed in longitudinal direction and vice versa. Figure 1 shows a re-entrant cellular auxetic exhibiting negative Poisson's ratio of about -0.4. A recent comprehensive review on auxetic metamaterials is available in Ref (K. K. Saxena, Das, and Calius 2016). In cellular materials, which can be engineered to have a wide range of negative values of Poisson's ratio, it can lead to considerable improvements in mechanical properties such as indentation resistance, acoustic shielding, wave steering, energy channeling and fracture toughness. Therefore, auxetic materials have potentially interesting properties for impact protection devices. For protection against impact, for example, it is necessary to develop a cellular structure which shows sufficient auxetic behavior, good strain sensitivity (response to small strains) and exhibits tunable stiffness as per the application requirement.

Recently, multimaterial additive manufacturing has shown capability to introduce material gradients in a 3D printed material. The use of multimaterial concept in additive manufacturing has been explored by different researchers in various disciplines. In the works of Rajkowski et al. (Rajkowski et al. 2009), multi-material milli-robot prototyping process was used to manufacture robot gripper and torsion hinge of two materials. The flexible polymeric material served as a hinge and the rigid polymeric material provided the desired stiffness to the remaining part. Stankovic et al. (Stankovic et al. 2015) developed an optimality criteria for finding optimality solutions of a multi-material lattice with fixed topology and truss cross section sizes using the empirical material measurements. The synthesis of multimaterial compliant mechanisms was presented in the work of Saxena et al (A. Saxena 2005). In the work of Rutz et al. (Rutz et al. 2015), multimaterial 3D bio-printing has been demonstrated towards creating functional tissues and the tunable bioink (prepared from mixing of other materials) offers an additional means to customize mechanical, chemical, physical and biological properties of printed structures. In the work of Kokkinis et al. (Kokkinis, Schaffner, and Studart 2015), 5D design space was proposed in 3D printing. Besides the three dimensions in 3D printing, two other dimensions that can control material properties were introduced. These were local control of composition and particle orientation. To control the particle orientation, low magnetic field was applied on deposited 3D printing inks (preloaded with magnetic platelets). Chan et al. (Chan et al. 2012) fabricated multi-material hydrogel cantilevers with varying stiffness using a 3D stereolithographic printer. Inspired from anisotropy in natural/biological systems, Oxman (Oxman, Tsai, and Firstenberg 2012) presented a digital anisotropy approach which can generate spatially controlled property gradients, thereby offering an additional way to control material properties. In an another work of Oxman (Oxman et al. 2014), a functional design of acoustic chaise inspired from Gemini was presented which implemented Stratasys's Connex technology. This technology can generate as many as 1200 material combinations using 3 base materials, thereby generating wide range of desired properties. Their objective was to design a multifunctional skin for Gemini that can act as integrated structural surface, an acoustical barrier and comfortable cushion. Wang et al. (Wang et al. 2016) demonstrated the possibility of fabrication of patient specific tissue mimicking phantoms using dual material 3D printed metamaterials.

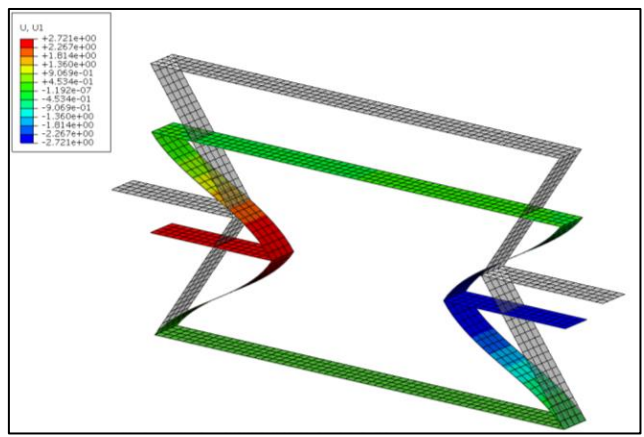

**Figure 1: Re-entrant hexagonal cell exhibiting negative Poisson's ratio (Abaqus Simulia 6.14-2).**

From the above literatures, it follows that multimaterial 3D printing has the capability of combining multiple materials in a single geometry which is a promising development in material fabrication. The multimateriality offers an additional dimension to control properties in available design space. When 3D printing technologies are combined with the optimization methods based on pixels and voxel based design, precise control on the distribution of objects can be achieved. It has been shown in the works of Hiller and Lipson (Hiller and Lipson 2010) that introducing material gradients in specific patterns at voxel level can give rise to auxetic behaviors (i.e. negative Poisson's ratio) in digital composites. In their work, digital material design was used. Aluminium voxels were embedded in a specific pattern in acrylic and negative Poisson's ratios were obtained. Recently, multi-material 3d printing with Stratasys' polyjet process has been used to create two material re-entrant cell designs (Wang et al. 2015). These two-material designs have been shown to exhibit some auxetic behavior in experiments but detailed mechanical analysis and understanding is still needed, which is necessary for practical applications. Even, there exist multi-material 3D printers which have capability to print more than 5 materials (Stratasys 2016). The idea of combining the properties of more than one type of material was also explored in metallic trusses so as to tune the negative Poisson's ratio. In a paper by Hu et al. (Hu and Silberschmidt 2013), multimaterial additive manufacturing was used to fabricate composite samples with re-entrant inclusions of different shape. The design was shown to exhibit tunable Poisson's ratio. In the works of Hughes et al. (Hughes, Marmier, and Evans 2010), a periodic truss with tunable auxetic properties was designed using the struts of three different types of crystal structure (FCC, BCC, and simple cubic) and different elastic moduli. They concluded that by tailoring the relative stiffness of the component beams within the structure, it is possible to design an auxetic truss structure with a specific Poisson's ratio, shear modulus, and tensile modulus. The concept of using more than one material in a cellular structure has also been used previously to tune other exotic properties such as negative thermal expansion (Hughes, Marmier, and Evans 2010; Lim 2005; Ng et al. 2016; Wei et al. 2016) and negative compressibility (Grima, Attard, and Gatt 2008). Thus, multimateriality serves as an extra dimension to design and control properties of 3D printed auxetics within a given design space.

Multimaterial cellular auxetics are no more limited to theoretical concept, but they can be realized in practice using 3D printing techniques (Stratasys 2016). Much attention has been given to tailor the negative Poisson's ratio of a cellular structure by manipulating its geometrical parameters and architecture. However, limited studies have focused on tuning the stiffness of the auxetic cellular structure of fixed geometrical parameters for enhancing its applications. In this work, multimaterial cellular auxetics were designed and modelled using finite element method. Four commercial 3D printing polymers were used to design the cell. The effect of material distribution and combination on the Poisson's ratio and cellular stiffness was studied.

## 2   MATHEMATICAL BACKGROUND OF THE CONCEPT

For a single material structure, the stiffness can be tuned by varying the shape, dimensions and material of a structure. However, if the geometry and dimensions are restricted; the stiffness can be tuned by exploiting the benefits of multimateriality. Multimaterial 3D printing offers the advantage of increasing anisotropic stiffness by using more than one material in a structure. Here we are concerned with developing a multimaterial auxetic cellular structure that has tunable stiffness properties which can be realized in practice using multimaterial 3D printing. In the subsequent sections, FE development of the concept is presented. In this section, a theoretical motivation of the concept is presented. The model of Tateno (Tateno 2016) is well suited for stiffness based designing of multi-material 3D printed mechanical parts and can be restated here for motivating the multimaterial concept in auxetic structures also. Consider that an auxetic re-entrant structure is composed of 6 multimaterial beams. To simplify the analysis, one single multimaterial beam is considered. To illustrate this concept, consider a beam of cellular structure as a cantilever beam composed of two materials with unit cross-section. Material 1 has lower stiffness as compared to material 2 as shown in Figure 2. Because of its structure the re-entrant cell can deform by both stretching/compressing and bending, thus we introduce a stiffness anisotropy parameter $Z$ which is given as ratio of stiffness of beam under axial deformation ($S_x$) and bending deformation ($S_y$):

$$Z = \frac{S_x}{S_y} \qquad 1$$

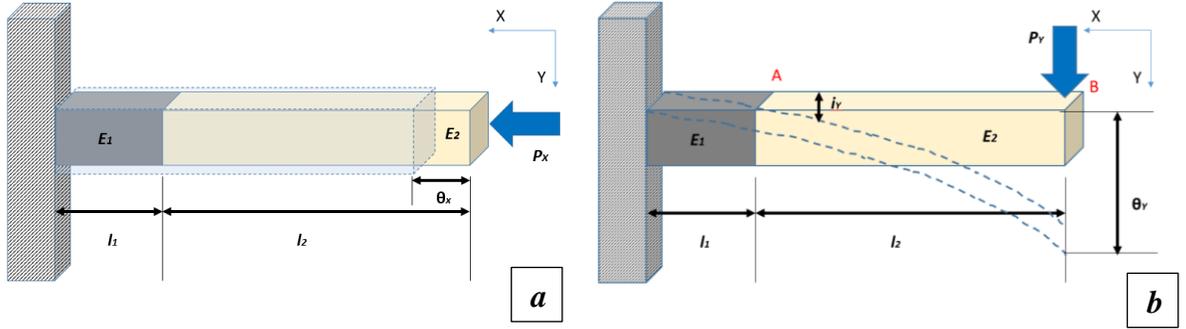

Figure 2: A two material beam with (a) axial point load (b) bending load at free end. The black color represents softer material and yellow color represents comparatively harder material.

For a single material cantilever beam of length l and unit cross section ($b = h = 1$) with a point load ($P$) in axial direction at free end, the axial displacement and stiffness in $X$ direction is given as :

$$\theta_x = \frac{Pl}{AE} = \frac{Pl}{E} \; (for \; A = 1) \qquad 2$$

$$S_x = \frac{Load}{Displacement} = \frac{E}{l} \qquad 3$$

For a single material cantilever beam of unit cross section with a point load on the free end, the maximum deflection is given as:

$$\theta_y = \frac{Pl^3}{3EI} = \frac{4Pl^3}{E} \qquad 4$$

Where I = 1/12 (Since $b = h = 1$) $\qquad$ 5

Thus, the bending stiffness in Y direction is given as:

$$S_y = \frac{Load}{Deflection} = \frac{E}{4l^3} \qquad 6$$

Thus for a single material cantilever beam, stiffness anisotropy parameter can be obtained using Eq 1:

$$Z_{single \; material \; beam} = 4l^2 \qquad 7$$

Now refer to Fig 2(a), the axial displacement (θx) due to load (Px) is given as:
$$\theta_x = \frac{P_x L_1}{E_1} + \frac{P_x L_2}{E_2} \quad (for\ A = 1) \tag{8}$$

As the axial displacement is mainly due to compression of softer material, therefore the displacement due to relatively harder material can be ignored to simplify the comparison.
Eq. 8 can be re-written as:
$$\theta_x \approx \frac{P_x L_1}{E_1} \tag{9}$$

Stiffness in X direction is given as:
$$S_x = \frac{E_1}{L_1} \tag{10}$$

The maximum deflection for a two material beam subjected to a point load at free end can be evaluated using Eq. 11:

$$\theta_Y = \theta_A + \theta_B + \theta_{A,B} \tag{11}$$
$$\theta_Y = \theta_A + \theta_B + i_Y \cdot l_2 \tag{12}$$

The maximum deflection is mainly due to the bending of softer material at free end. Thus, the term $\theta_B$ can be neglected. Therefore, the total displacement in Y direction can be written as:

$$\theta_Y = \theta_A + i_Y \cdot l_2 \tag{13}$$

$$\theta_Y = \frac{4P_y l^3}{E_1}\left(1 - \frac{3l_2}{2l} + \frac{l_2^3}{2l^3}\right) + \frac{6P_y l_1 l_2(l_1 + 2l_2)}{E_1} \tag{14}$$

The stiffness $(S_y)$ in Y direction can be evaluated by dividing load ($P_y$) with the total deflection in Y direction:

$$\theta_Y = \frac{E_1}{4l^3\left(1 - \frac{3l_2}{2l} + \frac{l_2^3}{2l^3}\right) + 6l_1 l_2(l_1 + 2l_2)} \tag{15}$$

Using Eq. 1, the stiffness anisotropy of two material beam can be evaluated as (Tateno 2016):
$$Z_{two\ material\ beam} = \frac{4l^3}{l_1}\left(1 - \frac{3l_2}{2l} + \frac{l_2^3}{2l^3}\right) + 6l_2(l_1 + 2l_2) \tag{16}$$

Comparing Eq. 7 and 8, it is evident that the stiffness anisotropy of two material beam is higher (for a short length of softer material) as compared to a single material beam. Thus, multimateriality can be used to tune the stiffness of a mechanical structure (Tateno 2016). Based on this background, we design multimaterial cellular auxetic structures with tunable stiffness properties in following sections.

## 3   CONCEPT AND DESIGN OF MULTIMATERIAL AUXETIC CELL

The classical re-entrant cell was chosen in this study because of its well established auxetic behavior. It has been already established that Poisson's ratio and stiffness of auxetic cellular structure depends on the reentrant angle, dimensions of struts and on the material properties of the constituent struts. However, within a given design space it is difficult to change the geometrical dimensions to tune these auxetic properties. Here we exploit the benefits of multimateriality to tune the auxetic properties within a given design space.

The multimaterial cellular design was conceptualized keeping into account the two-folded objective. The introduction of softer material (E~ 0.5 MPa) at the re-entrant corner will minimize the stiffness of the re-entrant corners in a 3D printed sample. Thus, the struts can do relative motion easily which can help in improving the auxetic behavior. The soft material serves as a hinge in a 3D printed sample (Fig. 3) to facilitate opening and closing of reentrant cell during loading.

The second objective was to tune the stiffness of auxetic cell. To achieve this, multimateriality concept was used. The cell was designed using four commercial 3D printing polymers with stiffness values of 2300, 1.4, 0.8 and 0.5 MPa. The stiffest base material (2300 MPa) constitutes 40% of the strut and remaining 60% is designed with three other softer materials in variable proportions. The cell was designed such that there exists a stiffness gradient as we move from re-entrant corner to the centre of strut of the re-entrant cell. The material near the re-entrant edge is softer so as to facilitate relative motion of strut and to minimize the stiffness of the re-entrant corners. As we move away from the edge, the material stiffness increases. Figure 3 shows the multimaterial re-entrant cell.

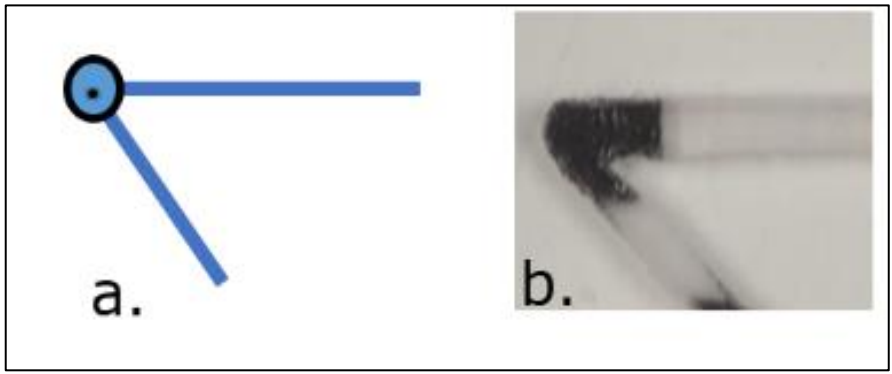

**Figure 3: (a) A reentrant corner with mechanical hinge. (b) Introduction of softer material at the reentrant corner in a 3D printed metamaterial. The softer material serves as a hinge.**

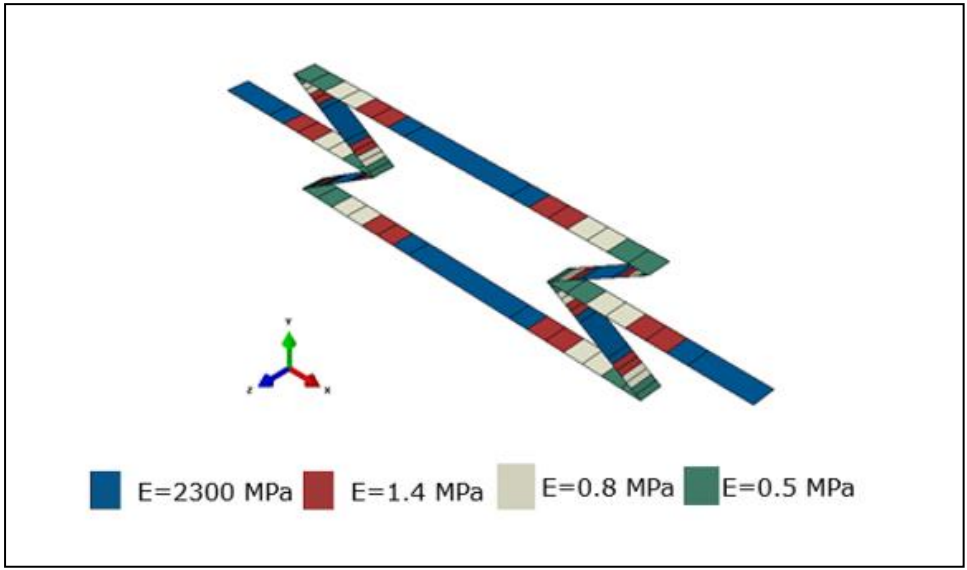

**Figure 4: A multimaterial reentrant cellular auxetic designed with four commercial 3D printing polymers.**

## 4. FINITE ELEMENT MODELLING

A finite element model was developed to evaluate the overall cellular stiffness and study the auxetic behavior of multi material re-entrant cellular structure. The representative volume element (RVE) was used to model the 35 x 10 mm cells as well as to facilitate its parameterization (Fig. 5). Periodic boundary conditions (Eq. 17) were imposed to represent a larger structure and different values of uniaxial compressive strains were then applied to the RVE.

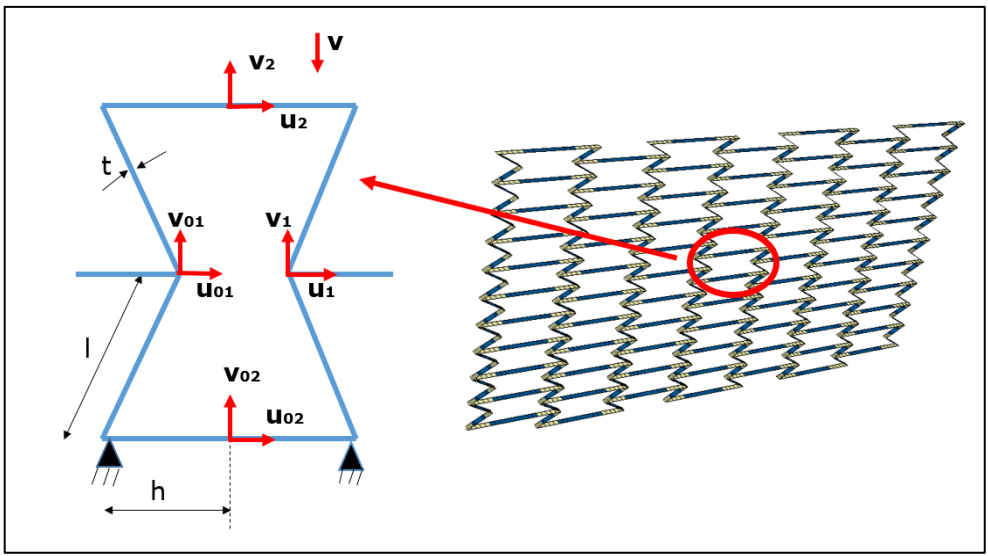

**Figure 5: Representative volume element (RVE) of the cellular structure with appropriate boundary conditions.**

$$\begin{aligned} v_1 &= v_{01} \\ u_2 &= u_{02} \\ v_2 - v_{02} &= v \end{aligned} \qquad 17$$

Three different elements i.e. solid, beam and shell were used to discretize the struts of the RVE. Truss elements (T3D2) were not used as they can carry only axial loads and this will not solve our purpose. Table 1 shows the element study using different types of solid elements, shell and beam elements. This study was done to determine the suitable element type and seed size for meshing before any of the primary simulations were carried out. To obtain the converged solution, the number of elements was changed by varying the global seed size for meshing and halving it with each subsequent trial. Figure 6 shows the comparison of model data using (solid (C3D8R), shell (S4R) and beam (B31)) for converged solution. The solid elements were too stiff and they give the lower values of Poisson's ratio which was against our consideration that multimateriality will lead to improvement in auxetic response. C3D8R solid element is highly compliant and is prone to hourglassing when there are very few elements in each strut along thickness. For C3D20 solid elements, computation time was very high and this is not desired when parametric studies have to be done. In case of meshing with beam elements, each cellular strut is represented using a 3 node beam element which involves bending. stretching, twisting and shearing deformation mechanisms. From preliminary simulations, it follows that using a beam element to model each strut of the cell is sufficient for convergence. However, it is also evident that the multimaterial cell is composed of short struts which cannot be properly represented using beam elements. S4R element makes use of uniformly reduced integration to reduce the effects of shear and membrane locking. It has capability to converge to shear flexible theory in case of thick shells and classical theory in case of thin shells. Thus, a single layer of shell elements was found suitable to model this cell in a computationally economic way and with the use of shell elements the effect of multimateriality or material choices can also be visualized conveniently. Simulations were carried out using FE package Abaqus 6.14-2.

A known displacement was applied in one direction and the transverse displacements were measured from relative nodal displacements. The Poisson's ratio was evaluated from Eq. 18.

$$\nu = -\frac{\epsilon_{trans}}{\epsilon_{long}} = -\frac{\epsilon_x}{\epsilon_y} \qquad 18$$

The effective Young's modulus of the RVE was evaluated from the ratio of averaged uniaxial stresses in the direction of applied strain and the strain in that particular direction as in Eq. 19.

$$E = \frac{\sigma_y}{\epsilon_y} \qquad 19$$

The multi-material cellular design as introduced in Fig. 4 was further parameterized by changing the distribution of the three softer materials. It should be noticed that the percentage of stiffest base material was kept same. Only the distribution of softer materials was changed. In total, four designs were simulated (as shown in Fig. 7) and their Poisson's ratio and stiffness values were evaluated.

**Table 1: Element study for simulating re-entrant multimaterial RVE.**

| Element | Seed size | No. of elements | Poisson's ratio |
|---|---|---|---|
| C3D8R | 1 | 780 | Hourglass effect |
|  | 0.5 | 5496 | -0.33 |
|  | 0.25 | 39648 | -0.33 |
| C3D8 | 1 | 780 | -0.32 |
|  | 0.5 | 5496 | -0.33 |
|  | 0.25 | 39648 | -0.33 |
| C3D20R (Quadratic) | 1 | 1024 | -0.34 |
|  | 0.5 | 6208 | -0.34 |
|  | 0.25 | 40880 | -0.34 |
| C3D20 | 1 | 1024 | -0.33 |
|  | 0.5 | 6208 | -0.33 |
|  | 0.25 | 40880 | -0.33 |
| C3D10 Tetrahedral (Quadratic) | 1 | 5947 | -0.33 |
|  | 0.5 | 38834 | -0.34 |
|  | 0.25 | 242211 | -0.34 |
| Linear wedge C3D6 | 1 | 1452 | -0.32 |
|  | 0.5 | 10496 | -0.33 |
|  | 0.25 | 75568 | -0.33 |
| S4R | 1 | 368 | -0.45 |
|  | 0.5 | 1240 | -0.45 |
|  | 0.25 | 4640 | -0.45 |
| S4 | 1 | 368 | -0.45 |
|  | 0.5 | 1240 | -0.45 |
|  | 0.25 | 4640 | -0.45 |
| B31 | 1 | 116 | -0.45 |
|  | 0.5 | 208 | -0.45 |
|  | 0.25 | 458 | -0.45 |

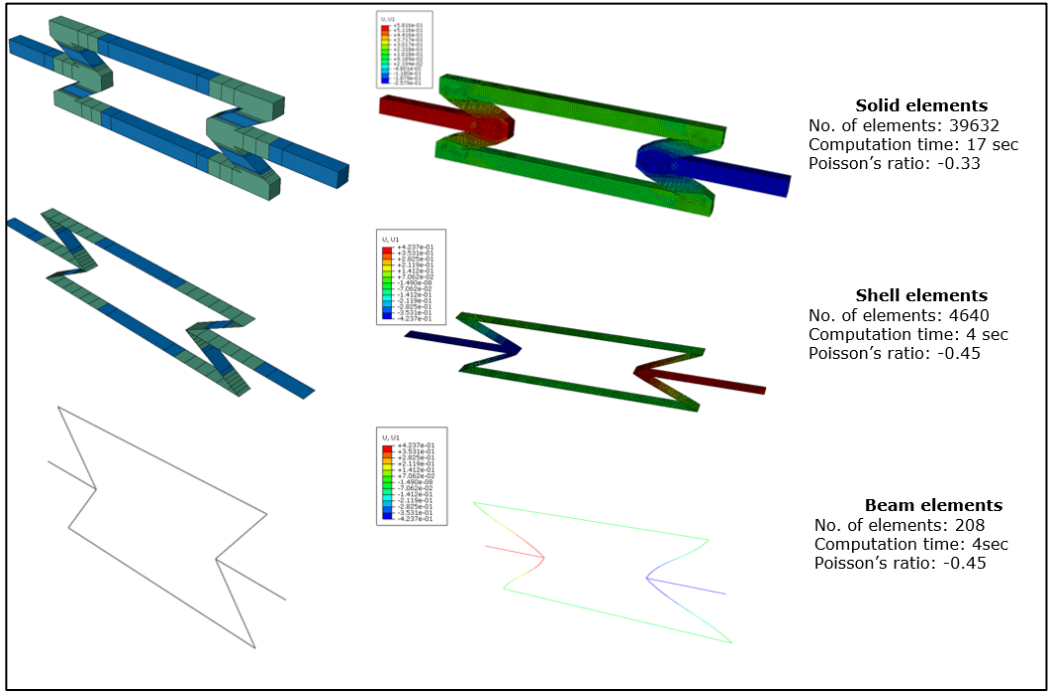

**Figure 6: Element study for the re-entrant multi-material RVE.**

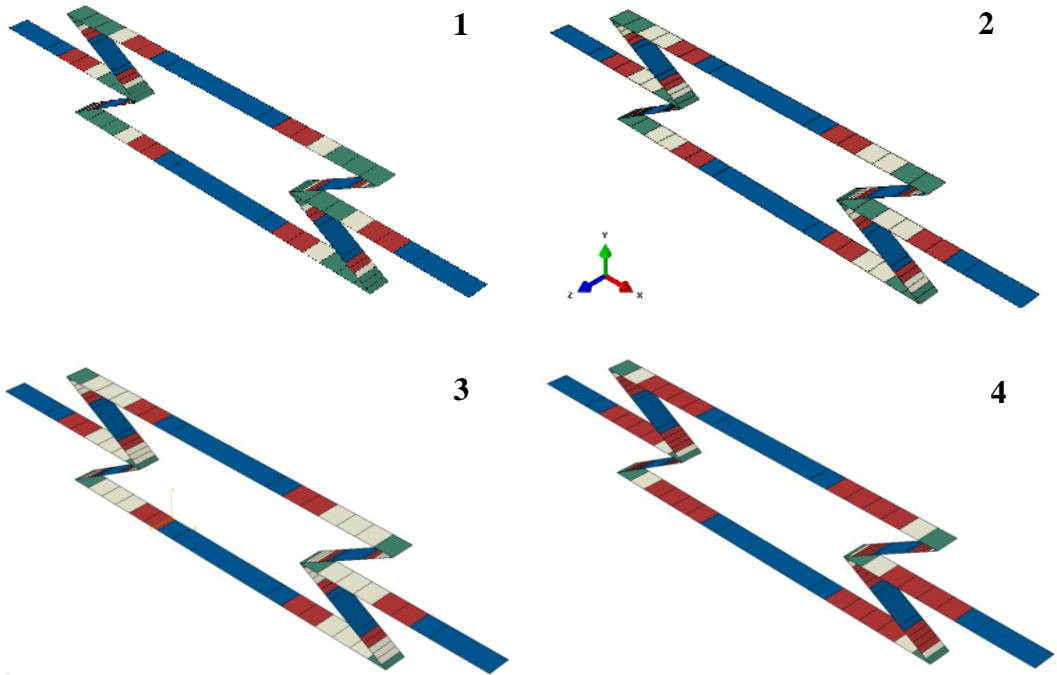

**Figure 7: The four conceptual designs of multimaterial reentrant auxetic cell.**

# 5   RESULTS AND DISCUSSION

## 5.1 EFFECT OF MULTIMATERIALITY ON POISSON'S RATIO

The Poisson's ratio of multimaterial cell was slightly higher than single material cell. Figure 8 shows the variation of Poisson's ratio for the multi-material cell (as in Fig. 4) with applied strain. It can be observed that at strains greater than 6% the Poisson's ratio becomes insensitive to applied strain. Thus, the Poisson's ratio of the multi-material re-entrant cell is influenced mainly by geometrical parameters. It is interesting to note that this multi-material cell is sensitive to very small strains, thereby increasing its suitability to applications which require response to very small strains (~1 %) such as wearable impact protection devices.

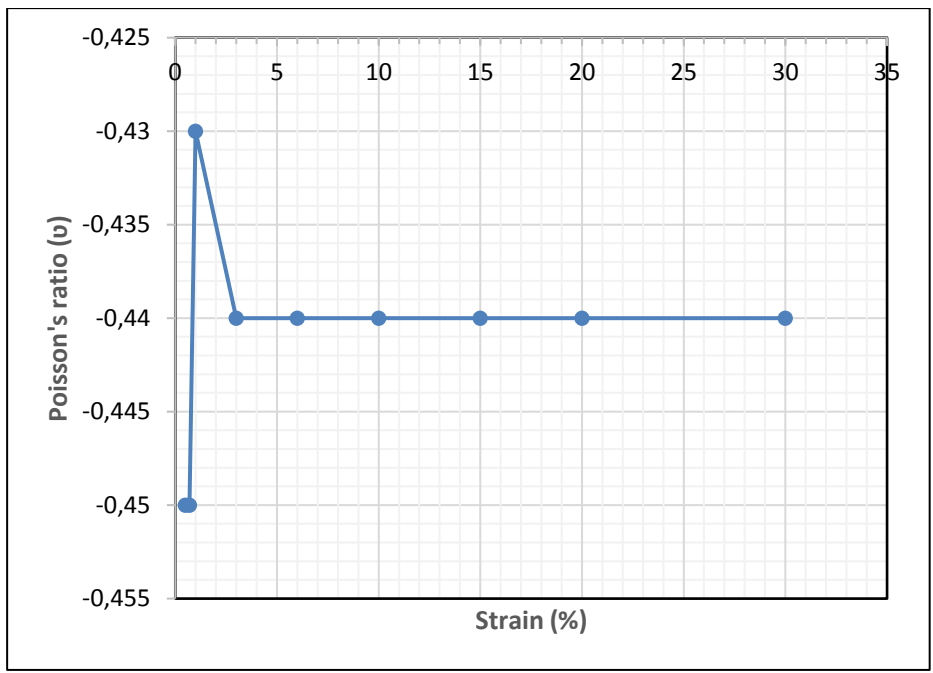

**Figure 8: Variation of negative Poisson's ratio with applied strain for the multi-material cellular design shown in Fig. 3.**

## 5.2 EFFECT OF MATERIAL CHOICES/DISTRIBUTION ON STIFFNESS

Figure 9 shows the results obtained after parametrizing the multimaterial cellular geometry. The parameter was the distribution and choice of softer material in the cellular strut. It can be observed from Fig. 9 that the distribution pattern of materials doesn't have significant effect on the Poisson's ratio. However, the overall stiffness of the cell changes considerably. Thus, material gradients in cellular struts affect Poisson's ratios marginally but they affect stiffness considerably. The choice and distribution of materials offer an additional way of tuning the cellular stiffness without compromising the negative Poisson's ratio. Effectively combining the knowledge of multimateriality and geometry, both auxetic properties and cellular stiffness can be tuned so as to enhance the applications of auxetics in 3D printed wearable impact protection devices in near future.

## 5.3 EFFECT OF BASE MATERIAL FRACTION ON CELLULAR STIFFNESS

Besides the material distribution, the fraction of stiffest base material in the cellular struts also affects the overall cellular stiffness. For evaluating this and facilitate parametrization, a two material design was used. Figure 10(a) shows a unit cell modelled using shell elements with dimensional parameters a = 30 mm and b = 10 mm. The parameter $l$ is taken as the length of inclined strut and $l'$ is the length of stiffest base material in the strut. The parameter $X$ (Eq. 20) is selected to describe the fraction of stiffest base material in the strut.

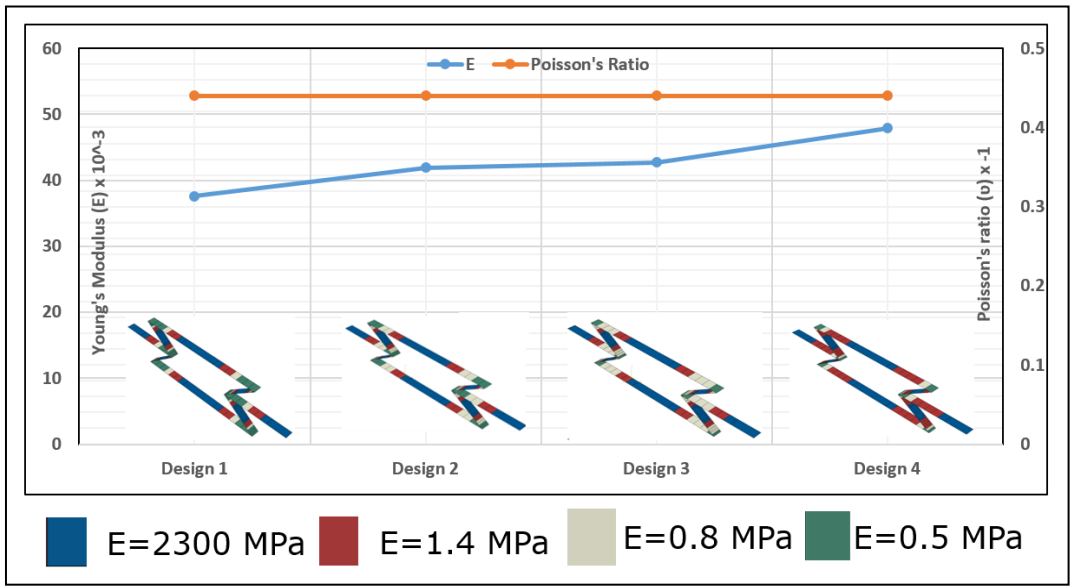

**Figure 9:** Dependence of Poisson's ratio and Young's Modulus on the material distribution in the cellular struts in a multi-material re-entrant cell.

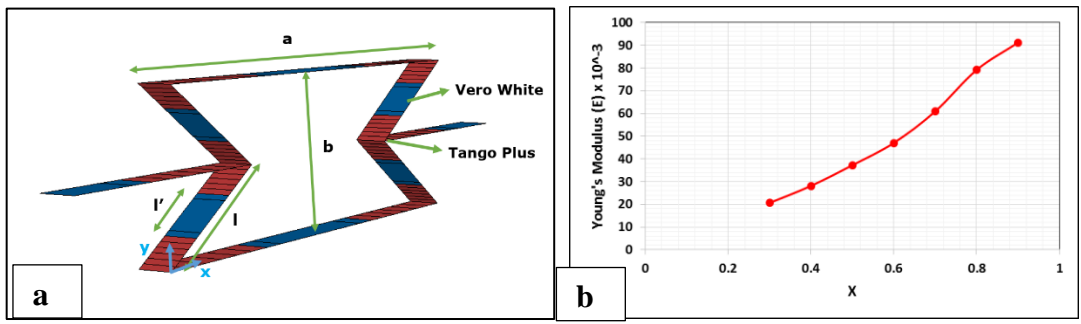

**Figure 10:** (a) A two material cell modelled using shell elements with details on image. (b). Variation of Young's Modulus of the cell with parameter $X$.

$$X = \frac{l'}{l} \qquad 20$$

Figure 10 (b) shows the variation of Young's Modulus of the cell with parameter $X$. It is clear that as the fraction of the stiffest material in the strut increases, the overall cellular stiffness also increases. The trend of variation of stiffness with $X$ is nonlinear. However, it can be seen that for an increase of base material fraction by 0.1, the stiffness increases by 10. Thus, the base material fraction also has an important influence on the overall cellular stiffness.

## 5.4 EFFECT OF MULTIMATERIALITY ON STRESS DISTRIBUTION

Figure 11 shows the stress distribution contours in the direction of applied load for a multi-material reentrant cellular structure. Since the shell elements have a local coordinate system; the s22 stress component is plotted for better visualization of stress distribution and interpretation. It can be observed that in case of a multi-material cell, the stresses also develop at the joints between materials although they are localized. There are very small and localized stresses at the re-entrant corners as the load is shared material joints in addition to the re-entrant corners. Thus, the re-entrant corners open up more easily during loading and result in improved auxetic behavior. Furthermore, the smaller as well as localized stresses at reentrant corners (except at material joints), permit easy translation and rotation of struts leading to structural flexibility which is an advantage for wearable applications.

It is of interest to determine the regions of stress singularities where the stress doesn't converges to a particular values. In order to determine the regions of stress singularities, a FE model using solid elements was developed as in Figure 12 (a) and (b). A 25% strain was applied and the Von Mises and Maximum Principal Stress was evaluated. It was observed that the material joint in all four chevron struts (Fig. 12 (c)) gave rise to stress singularity where stress was not found to converge. It was observed that both maximum principal stress and Von Mises stress were found to increase with the number of elements. Although the effect of singularity was localized and didn't affect overall model evaluations. As per St. Venant's principle, the effect of this singularity can be discarded as it is localized and doesn't affects the stresses away from this region.

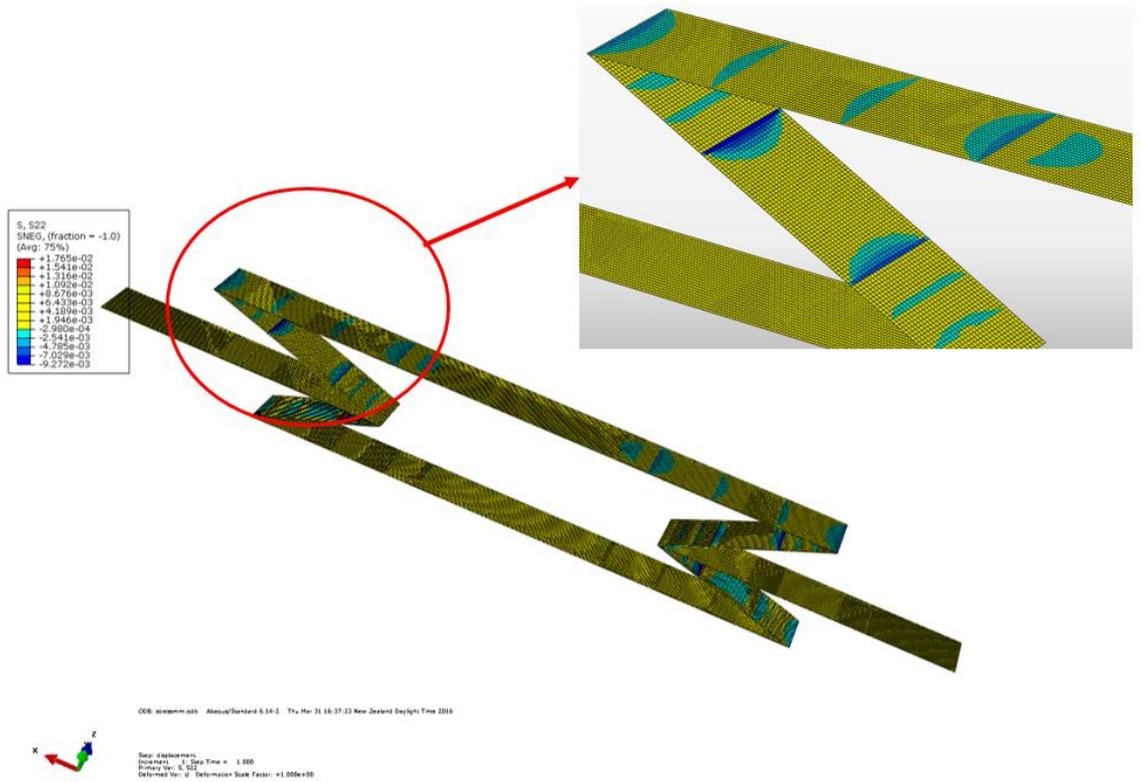

**Figure 11: Stress distribution on the struts of a re-entrant multimaterial cell.**

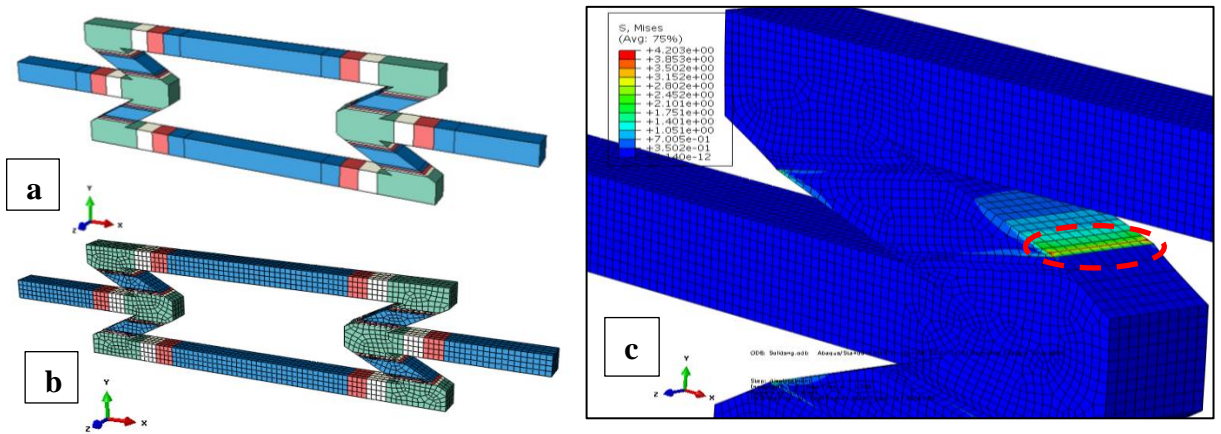

**Figure 12: (a) FE model of multi-material cell using solid elements (b) Meshed model and (c) Region of stress singularity in a re-entrant multimaterial cell (Red circle) for 25% applied strain.**

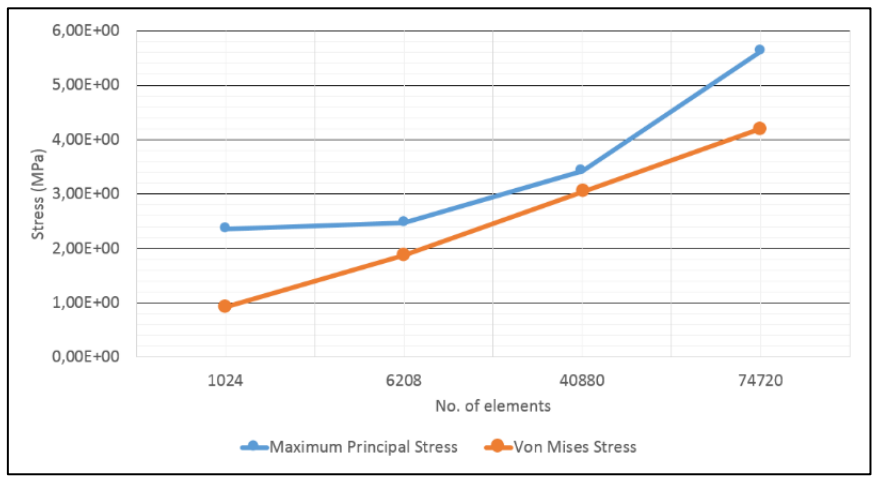

**Figure 13:** Plot of variation of maximum principal stress and maximum Von Mises stress with number of elements around the region of stress singularity in Figure 12(b) for applied strain of 25%.

## 6  FURTHER DISCUSSION

The results of this study can be useful in design of wearable impact protection devices which require three main properties in cellular materials; negative Poisson's ratio, a range of stiffness values depending upon the requirement and good strain sensitivity. The negative Poisson's ratios can be generated by manipulating the geometrical parameters. For a fixed set of geometrical parameters, the cellular stiffness can be tuned by introducing a material gradient in the cellular strut. Introducing soft material at the corners of the cell will also impart good strain sensitivity for protection against low velocity impacts. Our body imposes variable pressure distribution on the surfaces in contact , thus multimaterial cellular auxetics will help in minimizing pressure at high pressure points as in case of a footwear (Xu et al. 2015; Bickel et al. 2010) and also in case of sports helmets which require range of strength and crushing strains (Krzeminski et al. 2011; Naunheim et al. 2000). This concept of multi-material cell is useful in designing a cellular structure which shows wide range of stiffness values. These multimaterial cellular designs can also be used in objects of apparel. Besides the advantage of auxetic behavior, the combination of colors imparts aesthetic property and the stiffness of the cell can be tailored as per the need. Furthermore, the multi material cells can be used in animation objects which can give controlled and desired deformation, thereby making the animation characters more realistic (Schumacher et al. 2015). This work extends and validates the study performed on two material cellular auxetics as in the works of ref (Wang et al. 2015). This paper demonstrates our first capability in designing multimaterial cellular auxetic cells with tunable stiffness. Voxel level simulations can be useful in understanding the behavior of interfaces between the two materials fabricated by multimaterial 3D printing (Hiller and Lipson 2014).

## 7  3D PRINTED MULTIMATERIAL CELL

Figure 14 shows a 3D printed two-material cell for experimental investigations (K. K. Saxena, Calius, and Das 2016). The cell was printed using Stratasys' Polyjet process using two 3D printing polymers namely Vero White and Tango Plus. Figures 11 (b) and (c) show the deformed cell under manual compression and tension respectively. From manual stretching, the cell was found to have very good strain sensitivity with reduced corner stiffness. The experimental samples, camera based strain measurement and testing setup with appropriate load cell will help in experimental validation of the multi-material cellular approach in the near future.

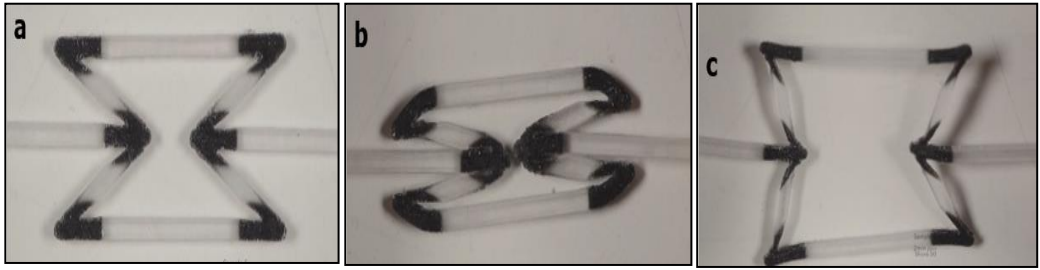

**Figure 14: A 3D printed two material (Vero White and Tango Plus) reentrant auxetic cell. (a) Unloaded (b) manually compressed (c) manually stretched.**

## 8   CONCLUSIONS

In this paper, the multi-material auxetic cell was designed and modelled using Finite element method. The effect of multimateriality on the auxetic behavior, strain sensitivity and cellular stiffness was analyzed using FE approach. The multi-material re-entrant cells exhibit high strain sensitivity i.e. they are sensitive to very low strains ~1%. This is desirable in impact protection devices, objects of apparel, animation character design. For constant geometrical parameters, multi-material auxetic cells can have wide range of cellular stiffness without compromising the auxetic response. This can be helpful in designing auxetic cellular network for specific applications which require variable stiffness. The material choice and distribution offers an additional dimension to tune the stiffness of auxetic cellular materials manufactured using 3D printing techniques. Further research can include investigation on other types of cellular configurations which can be explored for multimaterial concept. Voxel level simulations can be used for more in-depth study of material interfaces in multimaterial 3D printed auxetic structures. 3D printing of multimaterial cells can help in experimental validation. Precise strain measurement setup is required to validate the model data with the experimental results. The Young's modulus and strength of the printed cell can be correlated with the 3D printability.

**Data accessibility**

This paper has no linked data.

**Competing interests**

The authors declare no competing interests.

**Authors' contributions**

First author performed all the research work and wrote the manuscript. The second and third authors provided all the required guidance, revised this manuscript and the third author provided all the resources required for this work.

**Acknowledgements**

The authors would like to acknowledge the support from Callaghan Innovation for providing useful resources in this work.

**Funding statement**

The funding for this work was provided by University of Auckland and the required facilities were made available by Callaghan Innovation, Auckland, New Zealand.